\documentstyle[prl,aps,twocolumn]{revtex}
\def\Re{\, {\cal R}\mkern-3.1mu e\,}
\def\B.#1{{\bbox{#1}}}
\def\C.#1{{\cal{#1}}}
\def\BC.#1{\bbox{\cal{#1}}}
\def\sub.#1,#2 {_{\hbox{\tiny#1,{\scriptsize#2}}}}
\newcommand{\bec}[1]{\mbox{\boldmath $ #1$}}
\begin{document}
\title{Strong and weak clustering of   inertial particles in turbulent
flows}
\author{Tov Elperin$^1$ , Nathan Kleeorin$^1$, Victor
L'vov$^2$, Igor Rogachevskii$^1$ and Dmitry Sokoloff$^3$}
\address{$^1$ The Pearlstone Center for Aeronautical Engineering
Studies, Department of Mechanical Engineering,\\ Ben-Gurion University
of the Negev, Beer-Sheva 84105, P. O. Box 653, Israel}
\address{$^2$ Department of Chemical Physics, The Weizmann
Institute of Science, Rehovot 76100, Israel}
\address{$^3$Department of Physics, Moscow State University,
Moscow 117234, Russia}
\date{\today}
\maketitle
\begin{abstract}
We suggested a theory of clustering of inertial particles advected by
a turbulent velocity field caused by an instability of their spatial
distribution.  The reason of the {\em clustering instability} is a
combined effect of the particle inertia and finite correlation time of
the velocity field. The crucial parameter for the instability is a
size of the particles. The critical size is estimated for a {\em
strong clustering} (with a finite fraction of particles in clusters)
associated with the growth of the mean absolute value of the particles
number density and for a {\em weak clustering}  associated with the
growth of the second and higher moments. A nonlinear mechanism for a
saturation of the clustering instability (particles collisions in
the clusters) is suggested.  Applications of the analyzed effects to
the dynamics of aerosols and droplets in the turbulent atmosphere are
discussed. The critical size of atmospheric aerosols and droplets in
clustering is of the order of $(20 - 30)\mu$m, and a lower estimate of
the number of particles in a cluster is about hundreds.
\end{abstract}


\section{Introduction}

Formation and evolution of aerosols and droplets
inhomogeneities (clusters) are of fundamental significance in many
areas of environmental sciences, physics of the atmosphere and
meteorology ({\em e.g.} smog and fog formation, rain formation),
transport and mixing in industrial turbulent flows (like
spray drying, pulverized-coal-fired furnaces, cyclone dust separation,
abrasive water-jet cutting) and in turbulent combustion, see {\em
e.g.}~\cite{B2} and references therein.  Analysis of experimental data
shows that spatial distributions of droplets in clouds are strongly
inhomogeneous\cite{B99}.  Small-scale inhomogeneities in particles
distribution were observed also in laboratory turbulent
flows\cite{FKE94,HL00}.

It is well-known that turbulence results in a relaxation of
inhomogeneities of concentration due to turbulent diffusion\cite{B2},
whereas the opposite process, {\em e.g.}, a preferential concentration
({\em clustering}) of droplets and particles in turbulent fluid flow
still remains poorly understood.

In this Letter we suggest a theory of clustering of particles
and droplets in turbulent flows. The clusters of particles are formed
due to an instability of their spatial distribution suggested in
ref.~\cite{EKRI96} and caused by a combined effect of a particle
inertia and a finite velocity correlation time. Particles inside
turbulent eddies are carried out to the boundary regions between them
by inertial forces.  This mechanism of the preferential concentration
acts in all scales of turbulence, increasing toward
small scales. An opposite process, a relaxation of clusters is caused
by a scale-dependent turbulent diffusion. The turbulent
diffusion decreases towards to smaller scales. Therefore, the
clustering instability dominates in the Kolmogorov inner scale $\eta$,
which separates inertial and viscous scales. Exponential growth of the
number of particles in the clusters is saturated by their collisions.

\section{Qualitative analysis: instability growth rates for strong
and weak clustering}

For an estimate of cluster growth rates we use the equation for the
number density $ n(t,{\B.r}) $ of particles advected by a turbulent
velocity field $ {\B.u}(t,{\B.r}) $:
\begin{eqnarray}
\partial n / \partial t + \bec{\B.\nabla} \cdot(n {\B.v}) = D \Delta n
\,,
\label{T1}
\end{eqnarray}
where $ D $ is the coefficient of molecular (Brownian) diffusion.  Due
to inertia of particles their velocity ${\B.v}(t,{\B.r})\ne
{\B.u}(t,{\B.r})$. Equation (\ref{T1}) conserves the total number of
particles.  Equation for $ \Theta(t,{\B.r}) = n (t,{\B.r}) - \bar n $
follows from~(\ref{T1}):
\begin{equation}
\partial \Theta / \partial t + ({\B.v} \cdot \bec{\nabla}) \Theta = -
\Theta \,\mbox{div}\,\B.v + D \Delta \Theta \,,
\label{W25}
\end{equation}
where $ \bar n $ is the uniform mean number density of particles, and
the mean particles velocity is zero. We neglected the term $ \propto
\bar n\,\mbox{div}\,\B.v $ describing an effect of an external source
of fluctuations which is usually weaker than the effect of
self-excitation of fluctuations. In order to elucidate particles
clustering we analyze a role of different terms in eq.~(\ref{W25})
using a reference frame moving with a cluster velocity ${\B.V}_{\rm
cl}(t)$. The advective term, $ [({\B.v} - {\B.V}_{\rm cl})\cdot
\bec{\nabla}] \tilde \Theta$, causes turbulent diffusion inside the
cluster with the coefficient $D_{_{\rm T}} \sim \ell_{\rm cl} v_{\rm
cl}/3 ,$ where $ \ell_{\rm cl} $ is the characteristic size of a
cluster and $v_{\rm cl}$ is the turbulent velocity at the scale $
\ell_{\rm cl} $ and $\tilde \Theta(t,\B.r)$ are the fluctuations of
the number density in the comoving frame. The term $\propto
-\tilde\Theta \,\mbox{div}\,(\B.v-{\B.V}_{\rm cl}) $ can result in an
instability.  Indeed, neglecting diffusion we find the solution of
eq. (\ref{W25}): $\tilde \Theta (t,\B.r)\sim\tilde \Theta(0, \B.r)
\exp[-I(t)]$.~Here
$$ I(t)\equiv \int _0^t b(\tau) \,d \tau = \sum_{n=1}^{N} I_n
\,,\ I_n\equiv \int_{n\tau_v}^{(n+1)\tau_v}b(\tau) \,d \tau \,,
$$
and $b(\tau)=\mbox{div}[ {\B.v(\tau,\B.r)}-{\B.V}_{\rm cl}(\tau)]$ and
$N=t/\tau_v\gg 1$. We neglected a spatial dependence of $b(\tau,\B.r)$
inside the cluster and consider $b(\tau)$ as a random process with a
correlation time $\tau_{v}$ which is of the order of turnover time of
$\ell_{\rm cl}$-scale eddies, $\tau_{v}\sim \ell_{\rm cl}/ v_{\rm
cl}$. Now the integrals $ I_n$ can be considered as independent random
values and the sum $I(t)$ is estimated by the central limit theorem:
$I(t)\sim \sqrt{\langle \tau_{v} b^{2} \rangle_v \, t} \, \zeta ,$
where $\langle\dots \rangle_v$ is averaging over turbulent velocity
ensemble, $ \zeta $ is a Gaussian random variable with zero mean and
unit variance. Now, averaging over statistics of $\zeta$ by the
identity $ \langle \exp(k \zeta) \rangle_\zeta = \exp(k^{2}/ 2)$ (with
$k=\sqrt{\langle \tau_{v} b^{2} \rangle_v \, t} $ ), we estimate $
\langle |\Theta|^{q} \rangle \sim\langle |\Theta_{0}|^{q} \rangle \exp
(\gamma_q t)$ with the growth rate $ \gamma_q$ of the $q$-th moment
given by
\begin{equation}
\label{gq1}
\gamma_q \sim \case{1}{2} \langle \tau_{v} (\mbox{div} \, {\B.v}) ^{2}
\rangle_v  q^{2} - D_{_{\rm T}} q \,/ \ell_{\rm cl}^{2} \ .
\end{equation}
Here the turbulent diffusion inside the cluster is crudely taken into
account by the the term $\propto  D_{_{\rm T}}$.  Clearly, the
instability is caused by a nonzero value of $ \langle\tau_{v}
(\mbox{div} \, {\B.v})^{2} \rangle ,$ {\em i.e.,} by a {\em
compressibility of the particle velocity field} $\B.v(t,\B.r)$.

Compressibility of fluid velocity itself $\B.u(t,\B.r)$ (including
atmospheric turbulence) is often negligible and
$\mbox{div}\,\B.u=0$.
However, due to the effect of particle inertia their velocity
$\B.v(t,\B.r)$ does not coincide with $\B.u(t,\B.r)$ \cite{M87},
and a degree of compressibility of the field $\B.v(t,\B.r)$,
$\sigma_{v}$, defined by
\begin{equation}
  \label{eq:def-s}
  \sigma_{v} \equiv \langle [\mbox{div} \, {\B.v}]^2 \rangle_v /
  \langle |\B.\nabla \times {\B.v}|^{2} \rangle_v\,,
\end{equation}
may be of the order of $1$ \cite{EKRI96,EKR96,EKRS00}.  For inertial
particles $\mbox{div}\,\B.v \sim \tau_{\rm p} \Delta P / \rho $, where
$ \tau_{\rm p}$ is the Stokes time, $ \tau_{\rm p}= m_{\rm p} /6\pi
\rho \nu a =2 \rho_{\rm p} a^2 /9 \rho \nu $, $m_{\rm p}$, $\rho_{\rm
p}$ and $a$ are the mass, density and radius of particles,
respectively. The fluid flow parameters are: the viscosity $\nu$,
density $\rho$, pressure $P$, Reynolds number $\Re= L u_{_{\rm T}} /
\nu$ and the dissipative scale of turbulence $ \eta = L
{\Re}^{-3/4}$. Now we can evaluate:
\begin{eqnarray}
\sigma_v\simeq (\rho_{\rm p}/\rho)^2(a/\eta)^4\equiv (a/a_*)^4 \;
\label{W22}
\end{eqnarray}
(see \cite{EKRS00}), where $a_*$ is a characteristic radius  of
particles for which $\sigma_v=1$.  For water droplets in the
atmosphere $\rho_{\rm p}/\rho\simeq 10^3$ and $a_*\simeq \eta
/30$. For the typical value of $\eta\simeq 1$mm this yields $a_*\simeq
30\mu$m. On windy days when $\eta$ decreases, the value of $a_*$
correspondingly becomes smaller.

Then we estimate $ \langle \tau_v [\mbox{div} \,\B.v]^{2} \rangle_v$
as $ 2 \sigma_v/\tau_v$.  Taking the cluster size $ \ell_{\rm cl}$ of
the order of the inner scale of turbulence, $ \eta $, we have to
identify $\tau_v$ with a turnover time of eddies in the inner scale
$\eta$, $\tau_v \to \tau_\eta \equiv \eta/v_\eta\simeq (L/ u_{_{\rm
T}}) {\Re}^{-1/2}$. Thus, the growth rate $ \gamma_q$ may be evaluated
as
\begin{eqnarray}
\gamma_q\simeq \gamma _{\rm cl} \, q (q - q_{\rm cr}) \,,\quad \gamma
_{\rm cl} \sim \sigma_v/ \tau _\eta\,, \quad q_{\rm cr} \sim 1 / 3
\sigma_v \ .
\label{W27}
\end{eqnarray}
It is clear that moments with $q > q_{\rm cr}$ are unstable. It
follows from eqs.~(\ref{W22}) and (\ref{W27}) that this  happens when $
a>a_{q,\rm cr}$ where $a_{q,\rm cr}= a_{1,\rm cr}/ q^{1/4}$ is the
value of $a$ at which $q_{\rm cr}=q$. The largest value of $a_{q,\rm
cr}$ corresponds to the instability of the first moment, $ \langle
|\Theta|\rangle$: $a_{1,\rm cr} \sim 0.8 \,a_*$, $a_{2,\rm cr}
\approx 0.84 \,a_{1,\rm cr}$, $a_{3,\rm cr} \approx 0.76
\,a_{1,\rm cr}$, $a_{4\rm cr} \approx 0.71 \, a_{1,\rm cr}$, {\em etc.}

Note that if $ \langle |\Theta| \rangle $ grows in time then almost
all particles can be accumulated inside the clusters (if we neglect a
nonlinear saturation of such growth). We define this case as a {\em
strong clustering}. On the other hand, if $ q_{\rm cr} > 1 $
the first moment $ \langle |\Theta| \rangle $ does not grow and the
clusters contain a small fraction of the total number of particles.
This does not mean that the instability of higher moments is not
important. Thus, {\em e.g.}, probability of binary particles
collisions is proportional to the square of their density $\langle n
^2 \rangle=(\bar n)^2+ \langle |\Theta|^2 \rangle$. Therefore the
growth of the 2nd moment, $ \langle |\Theta|^{2} \rangle$, (which we
define as a {\em weak clustering}) results in that binary collisions
occur mainly between particles inside the clusters. The latter can be
important in coagulation of droplets in atmospheric clouds whereby the
collisions between droplets play a crucial role in a rain formation.
The growth of the $q$-th moment, $ \langle |\Theta|^q \rangle$,
results in that $q$-particles collisions occur mainly between
particles inside the cluster. The growth of the negative moments of
particles number density (possibly associated with formation of voids
and cellular structures) was discussed in \cite{BF01} (see also
\cite{SZ89,KS97}).

\section{Growth of the 2nd moment}

In a previous section we estimated the
growth rates of all moments $\langle |\Theta|^q \rangle$. Here we
present the results of a rigorous analysis for the evolution of the
two-point 2nd moment
\begin{equation}
  \label{eq:def-F}
  \Phi(t, \B.R) \equiv \langle \Theta(t, \B.r) \Theta(t, \B.r+\B.R )
  \rangle\ .
\end{equation}
In this analysis we used stochastic calculas ({\em e.g.,} Wiener path
integral representation of the solution of the Cauchy problem for
eq.~(\ref{T1}), Feynman-Kac formula and Cameron-Martin-Girsanov
theorem). We showed that a finite correlation time of a turbulent
velocity plays a crucial role for the clustering instability.
Notably, an equation for the second moment $\Phi(t, \B.R)$ of the
number density of inertial particles comprises spatial derivatives of
high orders due to a non-local nature of turbulent transport of
inertial particles in a random velocity field with a finite
correlation time \cite{EKRS00}. However, we found that at least for
two models of a random velocity field [(i) the random velocity with
Gaussian statistics of the integrals $\int _{0}^{t}\B.v(t',\bec{\xi})
\,d t'$ and $\int _{0}^{t} b(t',\bec{\xi}) \,d t'$; and (ii) the
Gaussian velocity field with a small yet finite correlation time] the
equation for $\Phi(t, \B.R)$ is a second-order partial differential
equation:
\begin{eqnarray}\label{WW6}
&&\partial \Phi / \partial t = \hat \C.L \, \Phi(t,\B.R) \,, \\
\nonumber
&&\hat \C.L = B(\B.R) + 2 {\B.U}(\B.R)\cdot \bec{\nabla}
+ \hat D_{\alpha \beta}(\B.R) \nabla_{\alpha} \nabla_{\beta} \ .
\end{eqnarray}
For the first model of the turbulent velocity field, {\em e.g.,}
the coefficients in eq.~(\ref{WW6}) are given by:
\begin{eqnarray}
 \label{eq:par}
B(\B.R) &\approx& 2 \int_{0}^{\infty} \langle b[0,\B.\xi(\B.r_1|0)]
b[\tau,\B.\xi(\B.r_2|\tau)] \rangle \,d \tau\,, \\ \nonumber
{\B.U}(\B.R) & \approx & - 2 \int_{0}^{\infty} \langle
{\B.v}[0,\B.\xi(\B.r_1|0)] b[\tau,\B.\xi(\B.r_2|\tau)] \rangle \,d\tau
\,, \\ \nonumber \tilde D_{\alpha \beta}^{^{^{\rm T}}}(\B.R) &\approx&
2 \int_{0}^{\infty} \langle v_{\alpha}[0,\B.\xi(\B.r_1|0)]
v_{\beta}[\tau,\B.\xi(\B.r_2|\tau)] \rangle \,d \tau \,,
\end{eqnarray}
where $\hat D_{\alpha \beta}(\B.R) = 2 D \delta_{\alpha\beta} +
D^{^{\rm T}}_{\alpha \beta}(\B.R) $ and $D_{\alpha \beta}^{^{^{\rm
T}}}(\B.R) = \tilde D_{\alpha \beta}^{^{^{\rm T}}}(0) - \tilde
D_{\alpha \beta}^{^{^{\rm T}}}(\B.R)$ is the scale-dependent tensor of
turbulent diffusion.

For the $ \delta $-correlated in time random Gaussian compressible
velocity field the operator $\hat \C.L$ is replaced by
$\hat \C.L_0 $ in the equation for
the second moment $ \Phi(t,\B.R) $, where
\begin{eqnarray}\label{W6}
\hat \C.L_0 &\equiv& B_0(\B.R) + 2 {\B.U}_0(\B.R) \cdot \bec{\nabla} +
\hat D_{\alpha \beta}(\B.R) \nabla_{\alpha} \nabla_{\beta} \,, \\
\label{WWW6} B_0(\B.R) &=& \nabla_{\alpha} \nabla_{\beta} \hat
D_{\alpha \beta}(\B.R) \,, \quad U_{0,\alpha}(\B.R) = \nabla_{\beta}
\hat D_{\alpha \beta}(\B.R) \,
\end{eqnarray}
(for details see \cite{EKRS00,EKRS99}). In the $ \delta $-correlated
in time velocity field the second moment $ \Phi(t,\B.R) $ can only
decay in spite of the compressibility of the velocity field. The
reason is that the differential operator $\hat \C.L_0 \equiv
\nabla_{\alpha} \nabla_{\beta} \hat D_{\alpha \beta}(\B.R) $ is
adjoint to the operator $\hat \C.L_0^\dag \equiv \hat D_{\alpha
\beta}(\B.R) \nabla_{\alpha} \nabla_{\beta}$ and their eigenvalues are
equal. The damping rate for the equation
\begin{eqnarray} \label{W1}
\partial \Phi / \partial t = \hat \C.L_0^\dag \, \Phi(t,\B.R) \,
\end{eqnarray}
has been found in ref.~\cite{EKR95} for a compressible isotropic
homogeneous turbulence in a dissipative range:
\begin{eqnarray}
\gamma_2 = - {(3 - \sigma_{_{\rm T}})^{2} \over
6\,\tau_\eta(1 + \sigma_{_{\rm T}}) (1 + 3 \sigma_{_{\rm T}})}
\,,
\label{W4}
\end{eqnarray}
where $\sigma_{_{\rm T}}$ is the degree of compressibility of the
tensor $D^{^{\rm T}}_{\alpha \beta}(\B.R)$. For
the $ \delta $-correlated in time incompressible velocity field
$(\sigma_{_{\rm T}} = 0)$ eq. (\ref{W1}) was derived in ref.~\cite{K68}.
Thus, for the model of turbulent advection with a delta
correlated in time velocity field the clustering instability of the 2nd
moment does not occur.

A general form of the turbulent diffusion tensor in a dissipative range is
given by
\begin{eqnarray} \label{W12}
 && D^{^{\rm T}}_{\alpha \beta}(\B.R)=  (C_{1} R^{2}
 \delta_{\alpha \beta} + C_{2} R_\alpha R_\beta )/\tau_ \eta \,,
\\  \nonumber
&&C_{1} = (4 + 2\sigma_{_{\rm T}})/3\, (1 + \sigma_{_{\rm T}})\,,
  \   C_{2} = (4\sigma_{_{\rm T}} - 2)/3\,( 1 + \sigma_{_{\rm T}}).
\end{eqnarray}
The parameter $\sigma_{_{\rm T}}$ is defined by analogy with
eq.~(\ref{eq:def-s}):
\begin{equation}
  \label{S}
  \sigma_{_{\rm T}}\equiv \frac{\B.\nabla \cdot \B.D_{_{\rm T}}\cdot
  \B.\nabla }{\B.\nabla \times \B.D_{_{\rm T}}\times \B.\nabla }=
  \frac{\nabla_\alpha \nabla_\beta D^{^{\rm T}}_{\alpha
  \beta}(\B.R)}{\nabla_\alpha\nabla_\beta D^{^{\rm T}} _{\alpha'
  \beta'}(\B.R) \epsilon_{\alpha\alpha'\gamma}
  \epsilon_{\beta\beta'\gamma} }\,,
\end{equation}
where $ \epsilon_{\alpha\beta\gamma}$ is the fully antisymmetric unit
tensor. By definitions~(\ref{eq:def-s}) and~(\ref{S}) $ \sigma_{_{\rm
T}}=\sigma_v$ in the case of $\delta$-correlated in time compressible
velocity field. Equations~(\ref{eq:par}) show that for a finite
correlation time identities~(\ref{WWW6}) are violated: $B(\B.R)\ne B_0
(\B.R)$, $\B.U(\B.R)\ne \B.U_0(\B.R)$.

Let us study the clustering instability. Equation~(\ref{WW6})
in a nondimensional form reads:
\begin{eqnarray}
{\partial \Phi \over \partial \tilde t} &=& { \Phi'' \over m(r)} +
\Big[{1 \over m(r)} + (U-C_{2}) r^{2} \Big]{2\,\Phi' \over r } + B
\Phi \,,
\nonumber \\
1/ m(r) &\equiv & (C_{1} + C_{2}) r ^{2} + 2/{\rm Sc} \,,
\label{W14}
\end{eqnarray}
where $\B.U \equiv U \B.R$ and $ {\rm Sc} = \nu / D $ is the Schmidt
number. For inertial particles $ {\rm Sc} \gg 1 $.
The nondimensional variables in eq.~(\ref{W14}) are
$r \equiv R/\eta$ and $\tilde t = t/\tau_\eta$, $B$ and $U$ are
measured in the units $\tau_\eta^{-1}$.  In a {\em molecular diffusion
region of scales} whereby $ r \ll {\rm Sc}^{-1/2}$, all terms $\propto
r^2$ (with $C_1\,,\ C_2$ and $U$) may be neglected. Then the solution
of eq.~(\ref{W14}) is given by $ \Phi(r) = (1 - \alpha r^{2})
\exp(\gamma_2 t)$, where $ \alpha = {\rm Sc} (B - \gamma_2 \tau_\eta) / 12 $
and $B > \gamma_2\tau_\eta$.  In a {\em turbulent diffusion region of
scales}, ${\rm Sc}^{-1/2} \ll r \ll 1$, the molecular diffusion term
$\propto 1/{\rm Sc}$ is negligible. Thus, the solution of eq. (\ref{W14})
in this region is $ \Phi(r) = A_{1} r ^{-\lambda} \exp(\gamma_2 t)$,
where $ \lambda = (C_{1} - C_{2} + 2 U \pm i C_3) / 2 (C_{1} + C
_{2})$, and $ C_3^2=4 (B-\gamma_2\tau_\eta)(C_1+C_2) - (C_{1} - C_{2}
+ 2 U)^{2}$.  Since the total number of particles is conserved in a closed
volume, $\quad \int_{0}^{\infty} r^{2} \Phi(r) \,d r =
0$. This implies that $ C_3^2 > 0$, and therefore $ \lambda $ is a
complex number. In addition, the correlation function $ \Phi(r) $ has
global maximum at $ r = 0 .$ This implies that $ C_{1} > C_{2}-
2U$. The latter condition, {\em e.g.,} for very small $U$ yields $
\sigma_{T} \leq 3$.  For $r \gg 1$ the solution for $\Phi(r)$ decays
sharply with $r$.  The growth rate $ \gamma_2$ of the second moment of
particles number density can be obtained by matching the correlation
function $ \Phi(r) $ and its first derivative $ \Phi'(r) $ at the
boundaries of the above regions, i.e., at the points $r =
{\rm Sc}^{-1/2}$ and $r = 1$. The matching yields $ C_3 / 2(C_{1} + C
_{2}) \approx 2 \pi / \ln {\rm Sc} $. Thus,
\begin{eqnarray}\label{W17}
\gamma_2 &=& {1 \over \tau_\eta (1 + 3 \sigma_{_{\rm T}})}
\biggl[{200 \sigma_{_{\rm U}} (\sigma_{_{\rm T}} - \sigma_{_{\rm U}})
\over 3 (1 + \sigma_{_{\rm U}})} - {(3 - \sigma_{_{\rm T}})^{2} \over 6 (1 +
\sigma_{_{\rm T}})}
\\ \nonumber
&&- {3 \pi^2 (1 + 3 \sigma_{_{\rm T}})^{2}
\over (1 + \sigma_{_{\rm T}}) \ln^2 {\rm Sc}} \biggr]
+ {20 (\sigma_{_{\rm B}} - \sigma_{_{\rm U}})
\over \tau_\eta (1 + \sigma_{_{\rm B}}) (1 + \sigma_{_{\rm U}})} \,,
\end{eqnarray}
where we introduced $\sigma_{_{\rm B}}$ and $\sigma_{_{\rm U}}$
defined by equations $ B = 20 \sigma_{_{\rm B}} / (1 + \sigma_{_{\rm
B}})$ and $ U = 20 \sigma_{_{\rm U}} / 3 (1 + \sigma_{_{\rm U}}) .$
Note that the parameters $ \sigma_{_{\rm B}} \approx \sigma_{_{\rm U}}
\sim \sigma_v $. For the $ \delta $-correlated in time random
compressible velocity field $ \sigma_{_{\rm B}} = \sigma_{_{\rm U}} =
\sigma_{_{\rm T}} = \sigma_v $. Note that eq.~(\ref{W4}) is written
for ${\rm Sc} \to \infty$.  Analysis of eq.~(\ref{W17}) shows that the
critical value of $\sigma_v$ required for the clustering instability
is $\sigma_{\rm cr} \approx (0.1 - 0.2)$. Notably, in the second model
of a random velocity field ({\em i.e.,} the Gaussian velocity field
with a small yet finite correlation time) the clustering instability
occurs when $\sigma_v > 0.2$. Equation~(\ref{W27}) also yields for the
threshold of the instability of the 2nd moment (at $q_{\rm cr} =2$) a
similar value $\sigma_{\rm cr}\sim 1/6$.

\section{Nonlinear effects}

The compressibility of the turbulent
 velocity field with a finite correlation time can cause the
 exponential growth of the moments of particles number density. This
 small-scale instability results in formation of strong
 inhomogeneities (clusters) in particles spatial distributions. The
 linear analysis does not allow to determine a mechanism of saturation
 of the clustering instability. As can be seen from eq.~(\ref{W17})
 molecular diffusion only depletes the growth rates of the clustering
 instability at the linear stage (contrary to the instability discussed
 in ref. \cite{BF01}). The clustering instability is saturated
 by nonlinear effects.

Consider now a mechanism of the nonlinear saturation of the
clustering instability using on the example of atmospheric turbulence
with characteristic parameters: $\eta\sim 1$mm, $\tau_\eta \sim (0.1 -
0.01)$s.  A momentum coupling of particles and turbulent fluid may be
essential when the mass loading is not small: $ m_{\rm p} n_{\rm
cl}\sim \rho$.  For this condition the kinetic energy of fluid $ \rho
\langle {\B.u}^{2} \rangle $ is of the order of the particles energy $
m_{\rm p} n_{\rm cl} \langle {\B.v}^{2} \rangle ,$ where $ |{\B.u}|
\sim |{\B.v}|$. This yields:
\begin{equation}
  \label{eq:est}
   n_{\rm cl} \sim a^{-3} (\rho/3 \rho_{\rm p}) \ .
\end{equation}
 For water droplets $ \rho_{\rm p} / \rho \sim 10^{3}$. Thus, for $ a
 = a_* \sim 30 \mu$m we obtain $ n_{\rm cl} \sim 10^{4} $ cm$^{-3}$
 and the total number of particles in the cluster of size $\eta$, $N_{\rm
 cl}\simeq \eta^3 n_{\rm cl} \sim 10$. These values may be considered
as a lower estimate for coupling effects of particles on the flow.

An actual mechanism of the nonlinear saturation of the clustering
instability is associated with particles collisions causing
particles effective pressure which prevents further grows of
concentration.  Particles collisions play essential role when during the
life-time of a cluster the total number of collisions is of the order of
number of particles in the cluster. The rate of collisions
$ J \sim n_{\rm cl} / \tau_\eta$ can be estimated as
$ J \sim 4 \pi a^{2} n_{\rm cl}^{2} |{\B.v}_{\rm rel}|$.
The relative velocity ${\B.v}_{\rm rel}$ of
colliding particles with different but comparable sizes can be
estimated as $ |{\B.v}_{\rm rel}| \sim \tau_{\rm p} | ({\B.u} \cdot
\bec{\nabla}) {\B.u}| \sim \tau_{\rm p} u_{\eta}^{2} / \eta$. Thus
the collisions in clusters may be essential for
\begin{equation}
  \label{eq:nc}
   n_{\rm cl} \sim a^{-3} (\eta/a)  (\rho/3 \rho_{\rm p}) \,,\quad
   \ell_{\rm s}  \sim a (3 a \rho_{\rm p}/\eta \rho)^{1/3}\,,
\end{equation}
 where $\ell_{\rm s}$ is a mean separation of particles in the
 cluster.  For the above parameters ($a=30\mu$m) $ n_{\rm cl} \sim
 3\times 10^{5}$cm$^{-3}$, $ \ell_{\rm s} \sim 5 \,a\approx 150 \mu$m
 and $ N_{\rm cl}\sim 300$. Note that the mean number density of
 droplets in clouds $\bar n$ is about $10^2 - 10
 ^3$cm$^{-3}$. Therefore the {\em clustering instability of droplets
 in the clouds increases their concentrations in the clusters by the
 orders of magnitude}.

In all our analysis we have neglected an effect of sedimentation of
particles in gravity field which is essential for particles of the
size $ a > 100 \mu$m. Taking $\ell_{\rm cl}\simeq \eta$ we assumed
implicitly that $\tau_{\rm p}< \tau_\eta$. This is valid (for the
atmospheric conditions) if $ a \leq 60 \mu$m.  Otherwise the cluster
size can be estimated as $\ell_{\rm cl}\simeq \eta (\tau_{\rm
p}/\tau_\eta)^{3/2}$.

Our estimates support the suggestion that {\em the clustering
instability serves as a preliminary stage for a coagulation of water
droplets in clouds leading to a rain formation}.

\section{Summary}

$\bullet$ We showed that the physical reason for the {\em clustering
instability} in spatial distribution of concentration of particles in
turbulent flows is a combined effect of the inertia of particles
leading to a compressibility of the particle velocity field
$\B.v(t,\B.r)$ and finite velocity correlation time.

$\bullet$ The clustering instability may result in a {\em strong
clustering} in which a finite fraction of particles are accumulated in
the clusters and a {\em weak clustering} when a finite fraction of
particle collisions occurs in the clusters.

$\bullet$ The crucial parameter for the clustering instability is a
radius of the particles $a$. The instability criterion is $a> a_{\rm
cr} \approx a_*$ for which $\langle (\mbox{div}\, \B.v)^2 \rangle
=\langle |\mbox{rot} \, \B.v|^2\rangle$. For the droplets in the
atmosphere $a_*\simeq 30\mu$m. The growth rate of the clustering
instability $\gamma_{\rm cl} \sim \tau_\eta ^{-1} (a/a_*)^4$, where
$\tau_\eta$ is the turnover time in the viscous scales of turbulence.

$\bullet$ We introduced a new concept of compressibility of the
turbulent diffusion tensor caused by a finite correlation time of an
incompressible velocity field. For this model of velocity field,
the field of Lagrangian trajectories is not divergence-free.

$\bullet$ We suggested a mechanism of saturation of the clustering
instability - {\em particle collisions in the clusters}.  An evaluated
nonlinear level of the saturation of the droplets number density in
clouds exceeds by the orders of magnitude their mean number density.

\acknowledgements
We have benefited from useful discussions with I. Procaccia.  This
work was partially supported by the German-Israeli Project Cooperation
administrated by the Federal Ministry of Education and Research, by
the Israel Science Foundation and by INTAS (Grant 00-0309). DS is
grateful to a special fund for visiting senior scientists of the
Faculty of Engineering of the Ben-Gurion University and to the Russian
Foundation for Basic Research for financial support under grant
01-02-16158.


\end{document}